# Coherence of Working Memory Study Between Deep Neural Network and Neurophysiology

Yurui Ming

*Abstract*—The auto feature extraction capability of deep neural networks (DNN) endows them the potentiality for analysing complicated electroencephalogram (EEG) data captured from brain functionality research. This work investigates the potential coherent correspondence between the region-of-interest (ROI) for DNN to explore, and ROI for conventional neurophysiological oriented methods to work with, exemplified in the case of working memory study. The attention mechanism induced by global average pooling (GAP) is applied to a public EEG dataset of working memory, to unveil these coherent ROIs via a classification problem. The result shows the alignment of ROIs from different research disciplines. This work asserts the confidence and promise of utilizing DNN for EEG data analysis, albeit in lack of the interpretation to network operations.

*Index Terms*—Attention Mechanism, Deep Neural Network (DNN), Electroencephalogram (EEG), Working Memory (WM).

## I. Introduction

The success of deep neural networks (DNN) in various fields draw attention of brain researchers to apply these models for electroencephalogram (EEG) data analysis, either to promote neuroscientific understanding or to facilitate brain-computer interface (BCI) [1-4]. Although it is not as strict as clinical requirement, the black-box operations of DNN still arise lots of concerns. For example, it is difficult to reach intuitive interpretations to the model behavior without knowing the underlying mechanism. Despite a long way to go for thorough understanding of the neural network dynamics [5-7], it is still obligatory to study the characteristic and to assert the feasibility of network models by linking and comparing the achievements with results from other methodologies.

The properties of EEG such as high temporal resolution, mobility, economy, etc. [8], confers its indispensable role for brain research. To blindly employ DNN to analyze EEG data might provide satisfying result catering to the application itself; however, loss of intuition might hinder the theoretic depth of the achievement. Compared with other DNN affiliated applications, certain brain research experiments are only conducted in restricted environment or idealized conditions due to practical constraints, and the results are demonstrated to community without real deployments of the models. This might lead the unintended consequences being buried into the learned models when utilizing neural networks, turning the overall research into hallucinations at a later stage [9-11]. Hence, to cross validate the results of DNN models in EEG data analysis by referencing knowledge from other disciplines is more critical than other applications.

In this work, we try to address part of these concerns by considering the implicit attention mechanism induced by class activation mapping via global average pooling (GAP) [12]. Actually, for certain fundamental brain research topics, such as working memory [13], there are already common recognition of neuronal basis underpinning this mechanism. It is regarded that prefrontal cortex (PFC) and hippocampus are actively involved in the functioning of working memory [14-16]. Therefore, it is expected that for working memory load test, when DNN is applied for harvesting EEG features automatically [2], the model does not switch absurdly among different areas over the scalp. Although there is still dispute over the characteristics of neuronal activities, such as discrete dynamics or sustained activities [17], the distinct among workload extents should only incur the network focus on approximately common areas.

Work in [12] demonstrates that by adoption of GAP, even trained via class level labels, the network can still exhibits localization ability, which can identify the discriminative image regions in a picture. This apparent simplicity offers a big advantage to verify some conjectures, such as for the case of working memory, that the network tends to look at similar areas with different activation strength. It is known that due to still limited understanding of working memory, to enforce network to explicitly explore specific regions might not be a good idea. Instead, network's intrinsic dynamics and behaviors, if it is in accordance with the assumption, is a strong evidence to the feasibility of utilizing DNN to analyze EEG data. And this is the purpose and potential contribution of our work in this paper.

## II. Methodology

As mentioned above, working memory is an indispensable component in many important cognitive functions and processes. Whether the elicited spiking pattern of neurons is continuously sustained or discretely dynamic, it is believed that high-order cortical areas, such as PFC are highly involved [18, 19]. Therefore, it is interesting to check where the network focus (the region-of-interest or ROI) to distinguish among the extent of workload during different workload exerted on working memory.

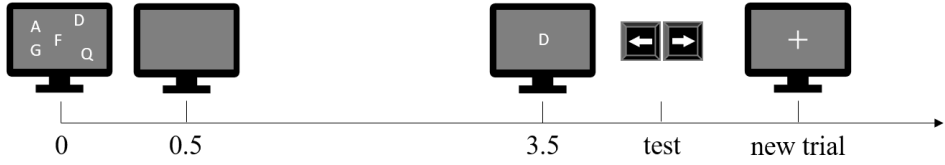

Fig. 1 Experiment paradigm for conducting the research.

## A. Experiment and Data

To investigate the behavior of network employed to analyze EEG data captured during a working memory capacity test [20], Fig. 1 shows the overall setup of the experiment. The paradigm of the working memory task for EEG dataset acquisition is as follows. A randomized letter set is displayed for participant at 0 second (s) relative to the beginning of current trial. It lasts for 0.5 s then fades away. At 3.5 s, a letter appears at the screen and subject needs to decide whether it belongs to the previous shown letter set or not on perceiving the letter. After the subject's action, a fixation icon launches the new trial.

The EEG signals are consecutively captured during the entire sessions which comprises of various trials, and the data preparation is same as in [2]. In detail, the EEG time series data lasting for 3.5 s are sliced into 7 non-overlapping segments. Then Fourier transform is applied to each segment to obtain the power spectrums up to 30Hz for each channel. According to literatures which reveal the effectiveness of respective EEG sub-bands [20, 21], three bands, i.e., theta (4-7Hz), alpha (8-13Hz), and beta (14-30Hz), are considered in this research. For each band of individual channel, the squared absolute values within the frequency band are added up to measure the contribution of the electrode source. At last, together with the corresponding EEG montage used in the experiment, the topographical representations are generated. Fig. 2 shows the overall procedures in an illustrative manner. Fig. 2(A) displays the EEG waveform data in the time domain; Fig. 2(B) shows each segment is converted into the frequency domain by FFT. The $x$-axis indicates the channels and the $y$-axis represents absolute values of frequency components; Fig. (C) presents EEG topography generated from frequency data by interpolation and extrapolation according to coordinates of the electrode placement.

## B. Network Architecture

The corresponding constructed DNN for analyzing EEG data to unveil network dynamics is in Fig. 3. First, a convolutional network (convnet) is applied to the non-overlapping segmented topographical data at each time step. Note the weights of the convnet are shared when processing each segment. Then the processed data are fed into a recurrent network for further computation. To better explore the spatial information of topographies, convolution is used inside the recurrent cell instead of conventional linear transformation plus non-linear activation. Hence, the subnetwork is abbreviated as recur-convnet. In detail, assuming at time step $t$, the input to the recur-convnet is $x_t$, the cell state is $c_t$, the hidden state is $h_t$, and the values of input gate, forget gate and output gate are $i_t$, $f_t$, $o_t$, respectively. The computation is governed by the following formulas:

$$\chi_t = \text{concat}(i_t, h_{t-1}) \qquad (1)$$

$$[i_t, f_t, o_t] = \text{conv}(\chi_t, w_g) \qquad (2)$$

$$c_t = \sigma(f_t) * c_{t-1} + \sigma(i_t) * \tan(\text{conv}(\chi_t, w_i)) \qquad (3)$$

$$h_t = \sigma(o_t) * \tan(c_t) \qquad (4)$$

$w$ represents weights of different network circuits in above formulas.

After the recurrence, only feature maps of the last step are considered. GAP is applied to these feature maps to obtain resulting weights before the final classification. These feature maps and weights are used to construct the heatmaps as in [12]. TABLE I details the configuration of the network structure.

## C. Heatmap Generation and Investigation

The complexity of the cognitive process and EEG data acquisition incur the non-intuition to the data itself, which consequently hinders the interpretation to the outcome of the network. To explicitly demonstrate the implicit attention capability of the network, we use a natural image set, which is intended to categorize dog and cat, to illustrate the ROI of network when performing classification. Because the designed network has a recurrent part, so for each dog or cat image, it repeats for 7 times in forming a sequential sample, to resemble the EEG topographical data. It means the layout of input data are exact between two datasets, just the dog and cat image set are natural for observation. Fig. 4 illustrates the network's behavior on the dog-vs-cat image set [22], which are natural to verify the effectiveness of the designed network.

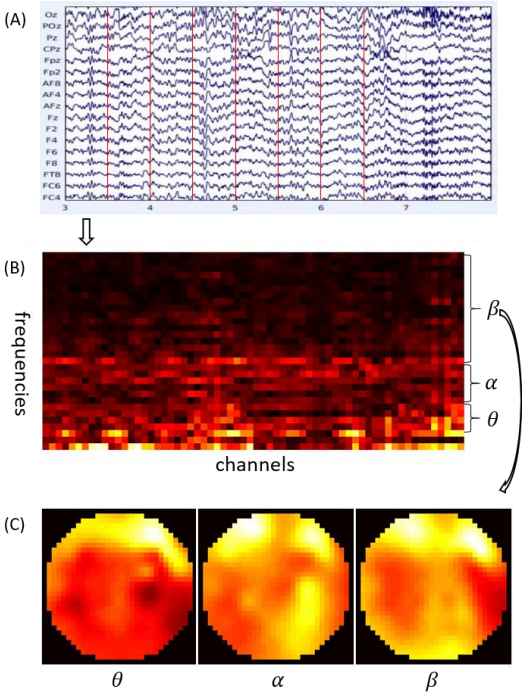

Fig. 2 Preparations of topographical EEG data in the spatial domain from waveform EEG data in the time domain.

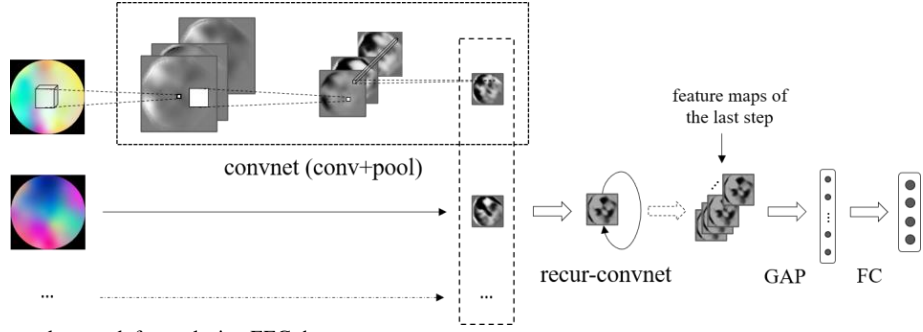

Fig. 3 Architecture of the neural network for analysing EEG data.

TABLE I NETWORK ARCHITECTURE

| Block | Layer | Filters | Size | Activation | Padding | Recurrence |
|---|---|---|---|---|---|---|
| ConvNet | Conv2D | 8 | (3, 3) | ReLU | SAME | |
| | Conv2D | 8 | (3, 3) | ReLU | SAME | |
| | Conv2D | 8 | (3, 3) | ReLU | SAME | |
| | Conv2D | 8 | (3, 3) | ReLU | SAME | |
| | AvgPool | | (2, 2) | | | - |
| | Conv2D | 16 | (3, 3) | ReLU | SAME | |
| | Conv2D | 16 | (3, 3) | ReLU | SAME | |
| | AvgPool | | (2, 2) | | | |
| | Conv2D | 32 | (3, 3) | ReLU | SAME | |
| | AvgPool | | (2, 2) | | | |
| Recur-ConvNet | Conv2D | 128 | (3, 3) | | SAME | 7 |
| GAP | AvgPool | | (8, 8) | | VALID | - |
| FC | Linear | | #class | Softmax | | - |

Based on the intuitive results achieved on the cat-vs-dog image set, the prepared EEG topographical data are processed by the network. We take a leave-one-subject-out test paradigm, which means each time we fix the data of one subject and use the data of all the remaining subjects to train the network. After training, the test data are fed into the network for prediction and simultaneously obtain the heatmap. To minimize the wrong prediction which can interfere the investigation, only samples with the right prediction are considered. The above procedure is repeated for all subjects with correctly predicted samples and corresponding heatmaps counting 2174 for analysis.

To investigate the properties of these heatmaps, for example, whether heatmaps corresponding to samples under different mind workload are separated or not, t-SNE is utilized to observe the distribution of these heatmaps in lower dimensional (D) space, such as in two-dimensional space [23]. Because the heatmaps are of high-dimensional data potentially spreading along certain manifold, instead of using Euclidean distance, the structural similarity index (SSIM) is considered here [24]. As inferred from [24], The SSIM measures the perceptual difference between two similar images and cannot judge which of the two is better. This is permittable with this research, since a specific heatmap is good or not is not confirmative, and only the collective heatmaps can propose certain conclusion. Therefore, as in [23], for heatmaps $h_i$ and $h_j$, the conditional probability is calculated as in (5):

$$p_{j|i} = \frac{\exp\left(\beta\left(1-\mathrm{ssim}(h_i,h_j)\right)^2\right)}{\sum_{j \neq k} \exp(\beta(1-\mathrm{ssim}(h_i,h_k))^2)} \quad (5)$$

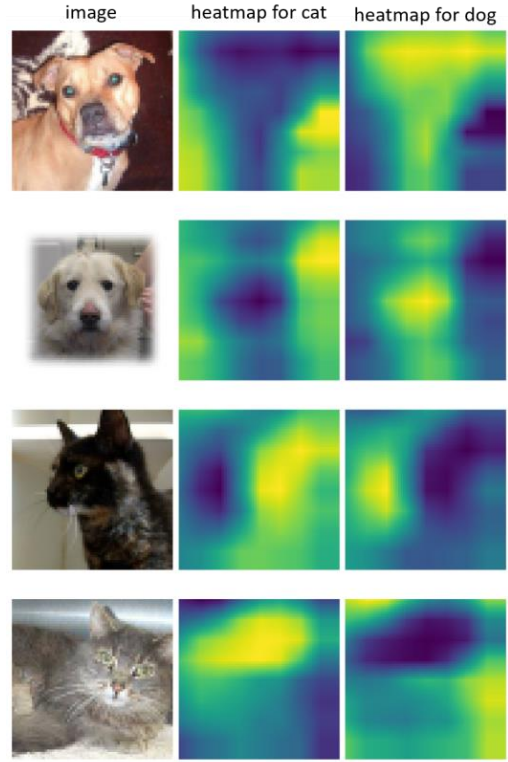

Fig. 4 Implicit attention capability of the designed network verified by the dog-vs-cat image set.

We take the perplexity of t-SNE as 20 to map the heatmaps from high dimensional space of 4096 into 2D plane. The algorithm is iterated for 1000 times.

## III. RESULTS

Fig. 5 illustrates the t-SNE result, from which it can help inspect the clustering attribute of all heatmaps. It is manifest that each class displays certain region preference but meantime spread the overall plane. However, no class obviously dominate specific region that clearly separates itself from other classes. This to some extent indicates that the network might work on some common parts of EEG topographies to distinguish the workload. These coherent parts of EEG topographies in turn indicate the activation of the corresponding brain regions. Fig. 5 indicates that for each class, the network will not focus differently at least apparently, which is optimistically supporting our conjecture. In addition, in Fig. 5, the clustering of class 2 is a little different from other classes. An explanation could be articulated as follows. For class 1, the cognition required to process the workload might be too simple to be distinguishable from the instinct; for class 4, the workload could be too high and complicated that requires sophisticated cognitive functions involved. Class 2 which represents appropriate workload is probably suitable for arousing cognition of working memory. This might explain its distinctiveness.

To further investigate the involved brain region, several heatmaps from different locations in Fig. 5 are displayed in Fig.

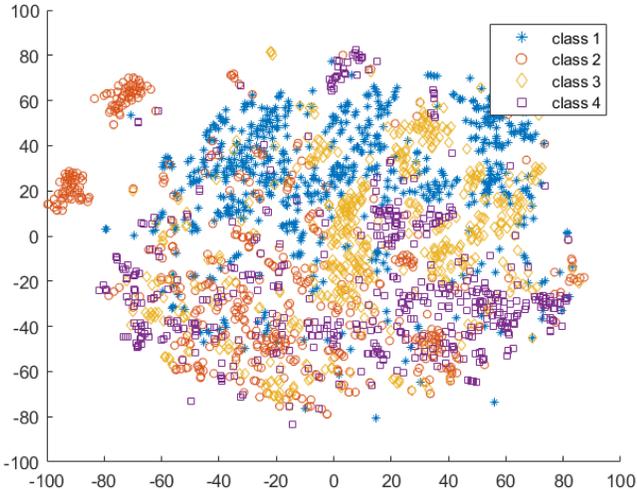

Fig. 5 Heatmap distribution under different workloads via t-SNE.

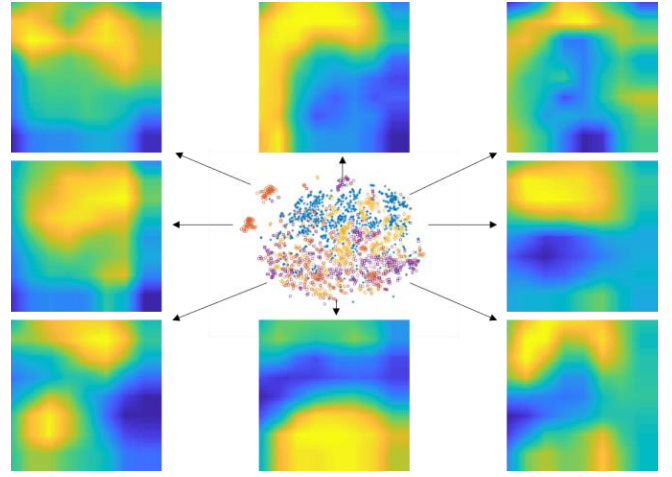

Fig. 6 Heatmaps from different locations of the t-SNE distribution.

6 to highlight this. It is interesting to notice that from these heatmaps, the PFC gets highly focused in all most all cases, although the activations might be on variance degrees. However, the result is sufficient to demonstrate the accordance with the conclusion of neurophysiological study as mentioned in the introduction part.

## IV. CONCLUSION

This paper investigated the potential conclusion coherence of working memory study between the approach from the DNN perspective and the method from the neurophysiological perspective. The results reveal the fact that the network overall looks into certain common area to distinguish the topographical EEG data under different workloads. And the brain areas tend to get focused are PFC area, which merits the conclusion from conventional neural physiological studies.

## REFERENCES

[1] Y. LeCun, Y. Bengio, and G. Hinton, "Deep learning," Nature, vol. 521, no. 7553, pp. 436-444, 2015/05/01 2015, doi: 10.1038/nature14539.
[2] P. Bashivan, I. Rish, M. Yeasin, and N. Codella, "Learning Representations from EEG with Deep Recurrent-Convolutional Neural Networks," arXiv.org, 2016.
[3] V. J. Lawhern, A. J. Solon, N. R. Waytowich, S. M. Gordon, C. P. Hung, and B. J. Lance, "EEGNet: a compact convolutional neural network for EEG-based brain-computer interfaces," (in eng), J Neural Eng, vol. 15, no. 5, p. 056013, Oct 2018, doi: 10.1088/1741-2552/aace8c.
[4] A. Vahid, M. Mückschel, S. Stober, A.-K. Stock, and C. Beste, "Applying deep learning to single-trial EEG data provides evidence for complementary theories on action control," Communications Biology, vol. 3, no. 1, p. 112, 2020/03/09 2020, doi: 10.1038/s42003-020-0846-z.
[5] M. D. Zeiler and R. Fergus, "Visualizing and understanding convolutional networks," in European conference on computer vision, 2014: Springer, pp. 818-833.
[6] J. Gilmer, J. Yosinski, and J. Sohl-Dickstein, "SVCCA: Singular Vector Canonical Correlation Analysis for Deep Learning Dynamics and Interpretability," arXiv.org, 2017.
[7] G. Montavon, W. Samek, and K.-R. Müller, "Methods for interpreting and understanding deep neural networks," Digital signal processing, vol. 73, no. C, pp. 1-15, 2018, doi: 10.1016/j.dsp.2017.10.011.
[8] Niedermeyer's electroencephalography : basic principles, clinical applications, and related fields, Sixth edition. ed. Philadelphia: Wolters Kluwer/Lippincott Williams & Wilkins Health, 2011.
[9] J. Shane. "When algorithms surprise us." https://aiweirdness.com/post/172894792687/when-algorithms-surprise-us (accessed 01/10, 2020).

[10] M. Minsky. "Embarrassing mistakes in perceptron research." https://www.webofstories.com/play/marvin.minsky/122 (accessed 01/10, 2020).

[11] M. Ribeiro, S. Singh, and C. Guestrin, ""Why Should I Trust You?": Explaining the Predictions of Any Classifier," arXiv.org, 2016.

[12] B. Zhou, A. Khosla, A. Lapedriza, A. Oliva, and A. Torralba, "Learning Deep Features for Discriminative Localization," in 2016 IEEE Conference on Computer Vision and Pattern Recognition (CVPR), 27-30 June 2016 2016, pp. 2921-2929, doi: 10.1109/CVPR.2016.319.

[13] B. Alan, "Working memory: looking back and looking forward," Nature Reviews Neuroscience, vol. 4, no. 10, p. 829, 2003, doi: 10.1038/nrn1201.

[14] J. O. Pernille, W. Helena, and K. Torkel, "Increased prefrontal and parietal activity after training of working memory," Nature Neuroscience, vol. 7, no. 1, p. 75, 2003, doi: 10.1038/nn1165.

[15] J. D. Murray, A. Bernacchia, N. A. Roy, C. Constantinidis, R. Romo, and X.-J. Wang, "Stable population coding for working memory coexists with heterogeneous neural dynamics in prefrontal cortex.(Report)," Proceedings of the National Academy of Sciences of the United States, vol. 114, no. 2, p. 394, 2017, doi: 10.1073/pnas.1619449114.

[16] E. Boran et al., "Persistent hippocampal neural firing and hippocampal-cortical coupling predict verbal working memory load," Science Advances, vol. 5, no. 3, p. eaav3687, 2019, doi: 10.1126/sciadv.aav3687.

[17] M. Lundqvist, J. Rose, P. Herman, Scott l. Brincat, Timothy j. Buschman, and Earl k. Miller, "Gamma and Beta Bursts Underlie Working Memory," Neuron (Cambridge, Mass.), vol. 90, no. 1, pp. 152-164, 2016, doi: 10.1016/j.neuron.2016.02.028.

[18] E. K. Miller, C. A. Erickson, and R. Desimone, "Neural mechanisms of visual working memory in prefrontal cortex of the macaque," The Journal of neuroscience : the official journal of the Society for Neuroscience, vol. 16, no. 16, pp. 5154-5167, 1996, doi: 10.1523/JNEUROSCI.16-16-05154.1996.

[19] M. G. Stokes, "'Activity-silent' working memory in prefrontal cortex: a dynamic coding framework," Trends in Cognitive Sciences, vol. 19, no. 7, p. 394, 2015.

[20] P. Bashivan, G. M. Bidelman, and M. Yeasin, "Spectrotemporal dynamics of the EEG during working memory encoding and maintenance predicts individual behavioral capacity," European Journal of Neuroscience, vol. 40, no. 12, pp. 3774-3784, 2014, doi: 10.1111/ejn.12749.

[21] O. Jensen and C. D. Tesche, "Frontal theta activity in humans increases with memory load in a working memory task," European Journal of Neuroscience, vol. 15, no. 8, pp. 1395-1399, 2002, doi: 10.1046/j.1460-9568.2002.01975.x.

[22] M. Research. "Dogs vs. Cats Create an algorithm to distinguish dogs from cats." https://www.kaggle.com/c/dogs-vs-cats/data (accessed 01/12, 2020).

[23] L. J. P. van der Maaten and G. E. Hinton, "Visualizing High-Dimensional Data Using t-SNE," Journal of machine learning research, vol. 9, no. nov, pp. 2579-2605, 2008.

[24] Z. Wang, A. C. Bovik, H. R. Sheikh, and E. P. Simoncelli, "Image quality assessment: from error visibility to structural similarity," IEEE transactions on image processing, vol. 13, no. 4, pp. 600-612, 2004.